\documentclass[12pt]{article} 
\usepackage{amssymb}
\usepackage{amsmath} 
\usepackage{epsfig}
\usepackage{graphicx}
\usepackage{harvard}
\citationstyle{dcu}
\usepackage{url}
\usepackage{color}
\usepackage{multirow}



\def\sobre#1#2{\lower 1ex \hbox{ $#1 \atop #2 $ } }

\newcommand{\bvarepsilon}{\mbox{\boldmath $\varepsilon$}}

\begin{document}

\def\E{{\mathbb E}}
\def\P{{\mathbb P}}
\def\R{{\mathbb R}}
\def\Z{{\mathbb Z}}
\def\V{{\mathbb V}}
\def\N{{\mathbb N}}
\def\NN{{\bf N}}
\def\X{{\cal X}}
\def\supp{{\rm Supp}\,}
\def\XX{{\bf X}}
\def\Y{{\bf Y}}
\def\G{{\cal G}}
\def\T{{\cal T}}
\def\cC{{\cal C}}
\def\C{{\bf C}}
\def\D{{\bf D}}
\def\U{{\bf U}}
\def\K{{\bf K}}
\def\H{{\bf H}}
\def\n{{\bf n}}
\def\m{{\bf m}}
\def\b{{\bf b}}
\def\g{{\bf g}}
\def\sqr{\vcenter{
 \hrule height.1mm
 \hbox{\vrule width.1mm height2.2mm\kern2.18mm\vrule width.1mm}
 \hrule height.1mm}} 
\def\square{\ifmmode\sqr\else{$\sqr$}\fi}
\def\one{{\bf 1}\hskip-.5mm}
\def\liml{\lim_{L\to\infty}}
\def\given{\ \vert \ }
\def\ze{{\zeta}}
\def\be{{\beta}}
\def\de{{\delta}}
\def\la{{\lambda}}
\def\ga{{\gamma}}
\def\th{{\theta}}
\def\proof{\noindent{\bf Proof. }}
\def\rate{{e^{- \beta|\ga|}}}
\def\A{{\bf A}}
\def\B{{\bf B}}
\def\C{{\bf C}}
\def\D{{\bf D}}
\def\MM{{\bf m}}
\def\lnt{{\Lambda^N}}
\def\S{{\mathcal{S}}}
\def\basis{{\rm Basis}\,}
\def\life{{\rm Life}\,}
\def\birth{{\rm Birth}\,}
\def\death{{\rm Death}\,}
\def\flag{{\rm Flag}\,}
\def\type{{\rm Type}\,}
\def\Cov{{\rm Cov}\,}
\def\btheta{{\boldsymbol{\theta}}}
\def\bphi{{\boldsymbol{\phi}}}
\def\bbeta{{\boldsymbol{\beta}}}
\def\bvep{{\boldsymbol{\varepsilon}}}
\def\bequ{\begin{equation}}
\def\eequ{\end{equation}}
\newcommand{\bequu}{\begin{equation*}}
\newcommand{\eequu}{\end{equation*}}
\newcommand{\ind}{\indent}
\newcommand{\matp}[1]{\mbox{\tiny \boldmath{$#1$}}}
	
\title{Aggregated functional data model for Near-Infrared Spectroscopy calibration and prediction} 
\author{ Ronaldo Dias, Nancy L. Garcia\thanks{Corresponding author: {\tt nancy@ime.unicamp.br}}, Guilherme Ludwig \vspace{-.4cm} \\ {\small University of Campinas, BRAZIL} \\ and Marley A. Saraiva \vspace{-.4cm} \\ {\small Federal University of Goi\'as, BRAZIL} }
 
\date{\today}

\maketitle

\abstract Calibration and prediction for NIR spectroscopy data are
performed based on a functional interpretation of the Beer-Lambert
formula. Considering that, for each chemical sample, the resulting
spectrum is a continuous curve obtained as the summation of overlapped
absorption spectra from each analyte plus a Gaussian error, we assume
that each individual spectrum can be expanded as a linear combination
of B-splines basis. Calibration is then performed using two procedures
for estimating the individual analytes curves: basis smoothing and
smoothing splines. Prediction is done by minimizing the square error
of prediction. To assess the variance of the predicted values, we use a
leave-one-out jackknife technique. Departures from the
standard error models are discussed through a simulation study, in
particular, how correlated errors impact on the calibration step and
consequently on the analytes' concentration prediction.  Finally, the
performance of our methodology is demonstrated through the analysis
of two publicly available datasets.   \endabstract

\section{Introduction} \label{sec:intro}

Different substances reflect light in a characteristic way. The
ultraviolet light has wavelength (in nanometers) in the range 1 -- 400
nm, visible light in 400--750 nm and infrared region in 750 to $10^6$
nm. The latter is subdivided into near-infrared (NIR) region:
750--2500 nm, mid-infrared (MIR) region: 2500--16,000 nm and
far-infrared (FIR): 16,000--$10^6$ nm. When materials are submitted to
different wavelengths, the overtones and combinations in the NIR band
will produce very complex patterns that characterize the constituents
of the sample. Usually the samples are submitted to different
wavelengths of intervals of less than 10 nm and the spectrum is a
curve, although it is only measured at discrete set of points. Hence,
the different characteristic curves of the constituents of the sample
overlap and give rise to a curve which is a sum of several curves
depending on the concentration of each substance.

There are several techniques to measure the constituents of a
sample. As expected, the very precise are expensive, so-called
analytical techniques, while others are relatively inexpensive at the
cost of precision, as in the case of NIR spectroscopy. Inexpensive
techniques usually require an ``instrument calibration'' step. Let us
begin with a quote regarding NIR spectroscopy: ``Near-infrared (NIR)
spectroscopy is a technique whose time has arrived. And for good
reason: it is unusually fast compared to other analytical techniques
(often taking less than 30 seconds), it is non-destructive, and as
often as not no sample preparation is required. It is also remarkably
versatile. If samples contain such bonds as C-H, N-H, or O-H, and if
the concentration of the analyte exceeds about 0.1\% of the total
composition, then it is very likely to yield acceptable answers --
even in the hands of relatively untrained personnel. The price to be
paid, however, is the preliminary work, typical of any chemometric
method. The instrument/computer system must be ``taught'' what is
important in the sample'' \cite{burns:ciurczak:2007}. Literature
is filled with practical applications of NIR spectroscopy such as food safety
testing, protein detection, pharmaceutical development, forensics, to
mention just a few. For introductory material on the
subject see \citeasnoun{shenk:westerhaus:1991},
\citeasnoun{davis:2000}, \citeasnoun{sies:2002},
\citeasnoun{brer:2003} and \citeasnoun{burns:ciurczak:2007}.

For practical reasons, it is only possible to measure the spectral
data at a finite number of wavelengths $t_1 < t_2 < \ldots <
t_T$. Often $T$ is in the range 100--200 or even more. For this
reason, these data are generally analyzed with multivariate data
analysis techniques \citeaffixed{brer:2003}{multiple linear regression
  (MLR), principal components regression (PCR) and partial least
  squares (PLS), see} which consider the spectrum as a set of $T$
different variables. In this case, the ordering of the wavelengths is
irrelevant and correlation among close points is not included in the
model. From an experimenter point of view it could be more informative
to describe the spectrum as function rather than as set of points,
hereby taking into account the physical background of the spectrum,
being the sum of absorption peaks for the different chemical
components, whose absorbance at wavelengths close to each other are
highly correlated. In addition, an important advantage of the
functional data analysis approach is to allow measurements to be taken
at different wavelengths for each calibration sample as well as for
the prediction samples a situation that cannot be treated with the usual 
multivariate methods.

The objective of this work is to propose a functional model to analyze
this data, based on an application of the Beer-Lambert law. The proposed 
framework is to assume that each absorbance function is a smooth curve 
that can be well approximated by a function belonging to a finite 
dimensional space ${\cal H}_K$ which
is spanned by $K$ (fixed) basis functions, in this case
B-splines. See, for example, \citeasnoun{silv:1986},
\citeasnoun{koop:ston:1991}, \citeasnoun{vida:1999},
\citeasnoun{dias:1999a}. Although this fact
might lead one to think that the non-parametric problem becomes a
parametric problem, one notices that the number of coefficients can
be as large as the number of observations, the smoothing will be
obtained by a penalization criteria with the penalizing factor chosen
by cross-validation \cite{wahb:1990}.

The idea of viewing the spectra as a continuous function was proposed
by \citeasnoun{alsb:1993}. Since then, several papers have been
written on the subject of non-parametric estimation of NIR
spectroscopy data, ranging from Neural Networks, Support Vector
Machine techniques, smoothing splines, kernel approximation among
others.  We refer the interested reader to the book of
\citeasnoun{ferraty:vieu:2006} and references therein.  Nevertheless,
we agree with the view of \citeasnoun{saye:kete:dari:2008} that the
potential of functional data analysis has not been grasped by most of
practitioners. Differently from the commonly used techniques, this
paper presents a non-parametric model-based approach that can be
easily implemented and analyzed and not only works for prediction but
also provides the individual calibration curves for each analyte. This
disagregation of the spectra leads to new developments such as jointly
calibration and prediction and outlier prediction to be addressed in
future work. 

In this work we analyze two artificial datasets, one with strongly
correlated measurement errors and another just weakly correlated to
demonstrate the impact of correlation in the calibration step of the
analysis. \citeasnoun{borg:thod:1992} cite the correlation as an
important issue in the analysis of spectra, since the fine sampling
usually results in large correlation between adjacent points in the
spectrum. Also, we apply the non-parametric analysis to two real
datasets. The first dataset consists of the absorbance curves of
polyaromatic hydrocarbons presented in the textbook  by
\citeasnoun{brer:2003} to illustrate multivariate calibration and
prediction techniques. This is an interesting sample to analyze since
it was designed to achieve an orthogonal design in the calibration
step. It was also analyzed, using a Bayesian perspective,
but only for the calibration step, by
\citeasnoun{dias:garcia:schmidt:2012}. The other one is the so-called
Tecator Data which is used by several authors to compare the techniques
in terms of prediction tool, for example in the works by
\citeasnoun{borg:thod:1992}, \citeasnoun{marx:etal:2009}, 
\citeasnoun{ferraty:vieu:2006}, \citeasnoun{perez:vieu:2006}.

This work is organized as follows: Section \ref{sec:BL} discusses the
suitability of the aggregated functional data model for the
spectroscopy data. Sections \ref{sec:calibra} and \ref{sec:predict}
show how the calibration and the prediction step, respectively, are
performed. We used simulated datasets with different degrees of
complexity (Section \ref{sec:simul}) as well as real data sets taken
from \citeasnoun{jorgensen:goegebeur:2007} (Section
\ref{sec:data}). All estimation procedures were made using the
software R, by \citeasnoun{Rproj}. The
computer code  is available from the authors upon request. 

\section{The model: Beer-Lambert law} \label{sec:BL}

A chemical sample is a compound of several {\it constituents}. A pure
chemical sample would be composed of only one constituent. Each
constituent of interest is called an {\it analyte}, the amount of the
analyte $\ell$ is called $y_{\ell}$, $\ell =1, \ldots, m$. A sample is
called {\it closed} if $y_1 + \ldots + y_m = 1$, i.e. when all
constituents in the compound are analyzed. It is very uncommon to
work with closed samples. In this work, we will assume that only a
subset of constituents is considered.

The term {\it spectral data} refers to the absorbances $W(t_1),
\ldots, W(t_T)$ measured at $T$ wavelengths $t_1 < t_2 < \ldots <
t_T$. The Beer-Lambert law for one sample considering $m$
analytes is the linear relationship between absorbance and
concentration of the analytes given by
\begin{equation}
 \label{eq:beer1}
 W(t) \,=\, \theta_0 (t) + \sum_{\ell=1}^m y_{\ell} \theta_{\ell}(t) + \epsilon(t),
\end{equation}

\noindent with the restriction
\begin{equation}
 \label{eq:beerrest}
 \sum_{\ell=1}^m \theta_{\ell}(t) = 0,
\end{equation}
for $t = t_1, \ldots, t_T$.  

In this work, we consider that expression (\ref{eq:beer1}) is true 
for all wavelengths in an appropriate interval $[A,B]$ with $\epsilon(t)$ 
being a Gaussian process with covariance function given by 
$\sigma(s,t) = {\rm Cov} (\epsilon(s),\epsilon(t))$. 

We shall restrict ourselves to expand the absorbance curves in the 
well-known cubic B-splines basis. That is, there exist a positive integer $K$
and a knot sequence $\xi_1 \leq \ldots \leq \xi_{K+4}$ such that 
\begin{equation}
\label{eq:e3}
\theta_{\ell}(t) \, = \, \sum_{k=1}^{K} \beta_{\ell,k} B_k(t), 
\end{equation}

\noindent where $B_k(t)$, $k=1, \ldots, K$ are cubic B-splines.  More precisely, 
the $k$-th B-spline of order $r$ for the knot sequence $\xi_1, \ldots, \xi_{K+r}$ 
is defined by
\[
B_{k}(t)=(\xi_{k+r}-\xi_{k})[\xi_{k},\ldots,\xi_{k+r}](\xi_{k}-t)_{+}^{r-1} \quad \mbox{for all} \quad t \in \R,
\]

\noindent where $[\xi_{k},\ldots,\xi_{r+k}](\xi_{k}-t)_{+}^{r-1}$ is
$r$th divided difference of the function $(\xi_k - t)_{+}^{r-1}$
evaluated at points $\xi_{k},\ldots,\xi_{r+k}$. For more details see
\citeasnoun{boor:1978}. Moreover, B-splines have an important
computational property, they are splines with smallest possible
support. In other words, B-splines are zero on a large
set. Furthermore, a stable evaluation of B-splines with the aid of a
recurrence relation is possible.

The model (\ref{eq:beer1}) can therefore be rewritten as 
\begin{equation}
 \label{eq:model_npf}
W(t)=\sum_{k=1}^K \left[ \beta_{0, k} + \sum_{\ell=1}^m y_{\ell} \beta_{\ell, k} \right] B_k(t)+\epsilon(t)
\end{equation}

\noindent where $B_k(t)$ corresponds to the $k$-th B-spline basis evaluated 
at $t$ and $\beta_{\ell,k}$ is the corresponding coefficient for the
$\ell$-th constituent. 

\section{Calibration Procedure} \label{sec:calibra}

First we are going to propose a procedure for {\it calibration} using
the approach of functional data analysis. The calibration problem can
be thought as a supervised learning procedure. Assume we are given $I$
samples of varying compositions. Therefore, our data consists of $I$
spectra measured by NIR instrument $W_i(t), i=1, \ldots, I$, observed
at a finite set of wavelengths. For simplicity, we are going to consider
that all of them were observed at the same wavelengths $t = t_1,
\ldots, t_T$, but this assumption is not necessary. Also, we are given a matrix
$Y$ with linear independent lines containing the concentrations
measured by a reference method $y_{i,\ell}$ for $i=1,\ldots,I$ and
$\ell = 1,\ldots, m$. We are going to assume that the curves $W( \cdot
)$ are smooth functions that are observed at discrete points and we
are going to use a functional form of the Beer-Lambert
equation. Calibration methods study how $Y$ varies with $W$. That is,
given the model (\ref{eq:beer1}), how do we estimate the functions
$\theta_{\ell}(t), \ell=0,1, \ldots, m$.

The interesting feature of (\ref{eq:model_npf}) restricted to the observed 
points is that it can be seen as a linear model,
\begin{equation*}
{\mathbf W} \, = \, {\mathbf X} \boldsymbol \beta + \bvarepsilon,
\end{equation*}

\noindent where $\boldsymbol \beta$ contains the parameters
$\beta_{\ell,k}$, $\ell =0, \ldots, m$, $k=1,\ldots, K$ to be
estimated, $\bvarepsilon$ represents the error vector, ${\mathbf W}$
is a stacked vector containing the $I$ observed spectra at the points
$t_1, \ldots, t_T$, ${\mathbf X}$ contains the suitable linear
coefficients, that can be written as $\mathbf{X} = \left(
  \mathbf{1}_{n \times 1} \mid \mathbf{Y} \right) \otimes \mathbf{B}$,
where $\otimes$ is the Kronecker matrix product and $\mathbf{B}$ an
ordinary B-spline design matrix. The restriction (\ref{eq:beerrest})
is included into the model, in the way suggested by
\citeasnoun{rams:silv:2005} to avoid a constrained optimization step: an
additional vector of zeros, with length $T$, is appended to the
observed $\mathbf{W}$, and the row vector $\left(0 \mid \mathbf{1}_{1
    \times m} \right)$ is appended beneath the matrix $\left(
  \mathbf{1}_{I \times 1} \mid \mathbf{Y} \right)$. The expanded model is
\begin{equation}
\left( \! \begin{array}{c}
\mathbf{W}_{IT \times 1} \\
\mathbf{0}_{T \times 1} \\
\end{array} 
\! \right) \, = \, \left( \left(\begin{array}{c|c}
\mathbf{1}_{I \times 1} & \mathbf{Y}_{I \times m} \\ 
\hline 0 & \mathbf{1}_{1 \times m} \end{array}\right) \otimes
\mathbf{B} \right) \boldsymbol \beta + \bvarepsilon \,.
\label{eq:fullmodel}
\end{equation}

In this work we are going to consider two procedures for estimation:
\begin{itemize}
\item[(a)] {\bf Basis smoothing:} In this case $K$ will be chosen in
  an {\it ad-hoc} manner suitable for estimating local features
  of the curves but also small enough to guarantee the required degree
  of smoothness. In this case, the coefficient vector
  $\boldsymbol\beta$ can be obtained as the ordinary least squares
  estimate, ignoring correlation which yields the explicit solution
\begin{equation}
\label{beta_chapeu}
\hat{ \boldsymbol \beta }=({{\mathbf X}^+}^\prime {\mathbf X}^+)^{-1}{{\mathbf X}^+}^\prime {\mathbf W}^+ ,
\end{equation}

in which ${\mathbf X}^+$ and ${\mathbf W}^+$ are ${\mathbf X}$ and
${\mathbf W}$ augmented with zeros as in equation
(\ref{eq:fullmodel}).
 
\item[(b)] {\bf Smoothing splines:} Here we are going to use every
  observation point as a knot, getting therefore a number of
  coefficients to be estimated as large as the number of
  observations. To achieve the desired smoothness we apply a penalty
  to the squared norm of the second derivative of the spline basis in
  the least squares problem
\[
\left ( {\mathbf W}^+ - {\mathbf X}^+ \boldsymbol \beta \right )^\prime \left( {\mathbf W}^+ - {\mathbf X}^+ \boldsymbol \beta \right) + \lambda \boldsymbol \beta^\prime \left(\mathbf{I}_{m \times m} \otimes \mathbf{R} \right) \boldsymbol \beta
\]

which yields the solution
\[
\hat{ \boldsymbol \beta }=({{\mathbf X}^+}^\prime {\mathbf X}^+ + \lambda \mathbf{I}_{m \times m} \otimes \mathbf{R})^{-1}{{\mathbf X}^+}^\prime {\mathbf W}^+,
\]

where $\mathbf{R}$ is a matrix with entries $\mathbf{R}_{i,j} =
\int_{-\infty}^\infty D^2 B_i(t) D^2 B_j(t) \mathrm{d}t $, and $D^2$
is the second order differential operator $\partial^2 / \partial
t^2$. The degree of smoothness will be controlled either by inspection
of the curves plots (which will be referred to as an ``eyeballing
method") or by minimization of the generalized cross-validation
criteria (GCV, see \citeasnoun{wahb:1990}). Notice particularly that
the minimum GCV for all analytes is not achieved by the minimization
of the observed aggregated curves, due to the triangle inequality
\[
\left\| D^2 \sum_{\ell=0}^m y_{i,\ell} \theta_{\ell}(t) \right\|^2 \leq \sum_{\ell=0}^m | y_{i,\ell} | \left\| D^2 \theta_{\ell}(t) \right\|^2,
\]

\noindent where $\left\| \cdot \right\|$ is the functional norm. This implies that the 
optimality of GCV based choices of $\lambda$ in ordinary smoothing problems cannot be 
extended to this calibration context, which leaves eyeballing as an equally valid choice.

\end{itemize}

\section{Prediction} \label{sec:predict}

Once we have the estimated analyte absorbance curves $\hat{\theta}_{\ell}(t)$,
$\ell = 0, \ldots, m$, we can proceed to perform {\it prediction}, the
second step of the analysis. At this point, we are given a new set of
$J$ spectra measured by the same NIR instrument, 
$W^\ast_j(t)$, $j = 1, \ldots, J$, all of them observed at the
wavelengths $t = t_1, \ldots, t_T$ (for the sake of simplicity we
maintain the same notation but the measured points for prediction can
be distinct from the ones used for calibration). Using this new set of
data, we want to predict now the new concentrations $y^\ast_{j,\ell}$
using the Beer-Lambert relationship
\begin{equation}
 \label{model_npf*}
 W^\ast_j(t)=\sum_{k=1}^K \left[ \beta_{0,k} + \sum_{\ell=1}^m \beta_{\ell,k}y^\ast_{j,\ell}\right] B_k(t) + \epsilon_j(t).
\end{equation}

A simple way to obtain estimates for $y_{j,\ell}^\ast$, $\ell=1,
\ldots, m$, $j=1, \ldots, J$ is to use the estimated spectra
$\hat{\theta}_\ell$, plug them into equation (\ref{model_npf*}) and
find which set of $\mathbf{y}^\ast_j$ minimizes the square error of
prediction, that is
\begin{equation}
\label{eq:pred}
\hat{\mathbf{y}}^\ast_j = \arg \min_{\mathbf{y}^\ast_j} \sum_{n=1}^T \sum_{\ell = 1}^m \left(W^\ast_j(t_n) - y^\ast_{j,\ell} \hat{\theta}_\ell (t_n) \right)^2.
\end{equation}

To assess the variance of such estimates, we use a leave-one-out
jackknife technique.
 
\begin{enumerate}
\item For $i=1,\ldots,I$, leave out all data related to $W_i(t)$ and
  find the estimate $\theta_{\ell}^{(-i)}(t)$. Then,

\item Use $\hat{\theta}_{\ell}^{(-i)}(t)$ to ``estimate'' $y_{i,\ell}$ as $\hat{y}^{(-i)}_{i,\ell}$.

\item Compare $y_{i,\ell}$ with $\hat{y}^{(-i)}_{i,\ell}$ using  
\[
S_{\ell}^{2}= \frac{1}{I} \sum_{i=1}^{I} \left ( y_{i,\ell} - \hat{y}^{(-i)}_{i,\ell} \right )^2 
\] 

\noindent to estimate the variance of $\hat{y}^{(-i)}_{i,\ell}$.

\item For any new curve $W_j^\ast(t)$ consider the confidence interval
  to be $\hat{y}^\ast_{j,\ell} \pm c S_\ell$ (since the estimators are
  normally distributed conditionally on the calibration sample).
\end{enumerate}

Notice that the normality of the estimators $\hat{Y}^\ast_{j,\ell}$ is
a consequence of assumption of the normality of the error processes
$\epsilon(t)$.  In our model, {\it cf.} expression (\ref{model_npf*}),
$W^\ast_j = \hat{\theta}_0 + A y^\ast_j + \epsilon^\ast$ where
$W^\ast_{j}$ is a $(T \times 1)$ vector containing the $j$th
prediction sample, $\hat{\theta}_0$ is a $(T \times 1)$ vector
containing the values of $\hat{\theta}_0(t_n)$ and $A$ is a $(T \times
m)$ matrix formed by $\hat{\theta}_c(t_n)$. Since the calibration
sample and the prediction sample are independent, the vectors
$W^\ast_{j}$, $\hat{\theta}_0$ and $A$ are independent. In this case,
the estimator $y^\ast_j$ given by (\ref{eq:pred}) is
\[
\hat{y}^\ast_j = (A^\prime A)^{-1} A^\prime ( W^\ast_{j}  - \hat{\theta}_0)
\]
which is normally distributed conditionally on the calibration
sample.
 
In order to compare the prediction performance of different approaches
we are going to use the Standard Error of Prediction (SEP), which is
the root mean square of the difference between the true and the
predicted content. The contribution of component $\ell$ is given by
\begin{equation}
  \label{eq:sep_ell}
  \mathrm{SEP}_\ell = \left (\frac{1}{J-1} \sum_{j=1}^J \left(
    y^\ast_{j,\ell} - \hat{y}^\ast_{j,\ell} \right) \right )^{1/2}
\end{equation}
and the overall SEP is given by
\begin{equation}
  \label{eq:sep}
  \mathrm{SEP} = \left( \frac{1}{mJ-1} \sum_{\ell=1}^{m} \sum_{j=1}^J \left(
    y^\ast_{j,\ell} - \hat{y}^\ast_{j,\ell} \right) \right )^{1/2}.
\end{equation}

\section{Artificial Datasets} \label{sec:simul}

We conducted a simulation study to assess the performance of the
estimators proposed in Sections \ref{sec:calibra} and
\ref{sec:predict}.  The algorithm was implemented in  {\tt R} language
\cite{Rproj} and is available upon request. 

\begin{figure}
\vspace{5cm}
	\centering
(a)		\includegraphics[angle=-90,width=16cm,totalheight=6cm]{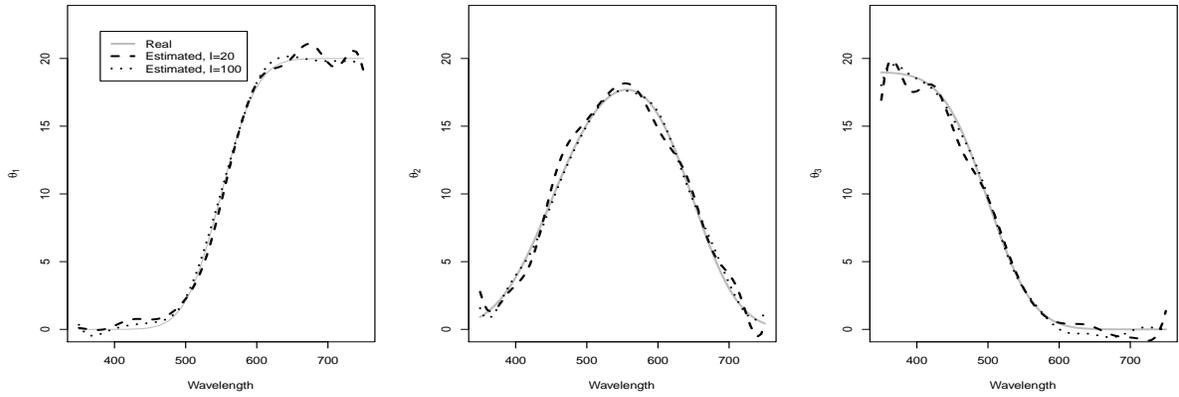} \\
(b)		\includegraphics[angle=-90,width=16cm,totalheight=6cm]{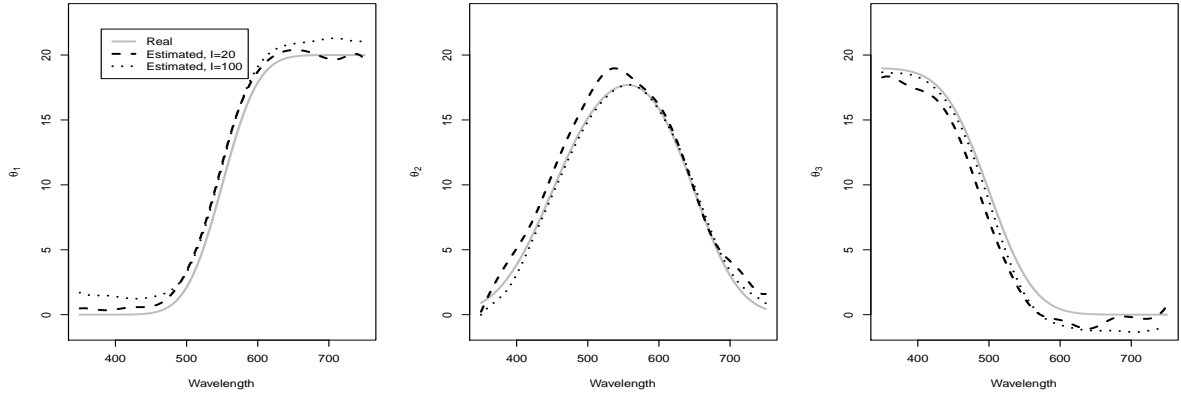} 
\caption{Estimated absorbance spectra, $\hat{\theta}_{1}(t),
  \hat{\theta}_{2}(t)$ and $\hat{\theta}_{3}(t)$, for (a) weakly
  correlated scenario and (b) strongly correlated scenario, using
  fixed $K$ basis.The dashed lines represent the spectra obtained when the 
 number of sampled curves is $I=20$, whereas the dotted lines represent
 the estimated spectra when $I=100$.}
\label{fig:simula5FittedThetaComparaI_new}
\end{figure}

The wavelengths are arbitrarily ranging from 350 to
750 units, with sample points taken every 5 units. Two samples were
simulated, one with $I=20$ curves and another including the 20
original observations and 80 additional ones, for a total $I =
100$. The samples were composed of $m=3$ analytes, with absorbances
shown in Figure \ref{fig:simula5FittedThetaComparaI_new},  
and concentrations $y_{i,1}, y_{i,2}, y_{i,3}$ randomly generated 
with a three-dimensional Dirichlet standard distribution 
($\alpha=1$ for all dimensions). The concentrations were generated
once and fixed afterwards, so they're accordingly treated as constants. 

Two scenarios were considered for the covariance function of each constituent,
$\sigma_i(s,t) = \sigma^2 e^{-\phi|t-s|}$, one with 
almost independent covariance structure for the analytes
($\phi=0.5$) and another one with
$\phi = 0.002$, resulting in a very strong auto-correlation for the
process. The variance was set to $\sigma^2=4$ for both cases. 

First we consider the results of a single experiment. The 
estimated analytes absorbance spectra $\hat{\theta}_{\ell}(t)$,
and their true equivalents, are shown in Figures
\ref{fig:simula5FittedThetaComparaI_new} (a) and (b).
The mean residual sum of squares 
for the weakly correlated data is $3.88$ when $I=20$ and $4.12$ 
when $I=100$; for the strongly correlated data we have $2.61$ when 
$I=20$, and $3.53$ when $I=100$. 

\begin{center}
[Figure \ref{fig:simula5FittedThetaComparaI_new} here] 
\end{center}

For the prediction step, the estimated standard deviations for the
concentration estimators $\hat{y}^\ast_1$, $\hat{y}^\ast_2$ and
$\hat{y}^\ast_3$, found via the leave-one-out estimation technique,
are shown in Table \ref{tab:sd}. We can see that in this scenario
the standard deviations for the strongly correlated data
range from approximately 2 to 4 times the standard deviations of the
weakly correlated data.

\begin{center}
[Table \ref{tab:sd} here]
\end{center}

The variance of the $\hat{\mathbf{Y}}^\ast$ estimators, in the
presence of strongly correlated errors, could be reduced if a
generalized least squares method is implemented during the calibration
step, as proposed by
\citeasnoun{dias:garcia:martarelli:2009}. However, the high
dimensionality of the problem and the fact that there is no
replication of the data results in poor aggregated covariance
estimates, with sample covariance matrices having high condition
numbers and therefore often being numerically singular. To overcome
this problem, we model the covariance function as suggested by
\citeasnoun{dias:garcia:schmidt:2012}, we assumed that the covariance
of each $\epsilon_i(t)$ in model (\ref{eq:model_npf}) is homogeneous
and given by
\[
\Cov( \epsilon_i(s), \epsilon_i(t)) = \Sigma_i (s,t) = \sum_{\ell=1}^m y_{i,\ell}^2 \sigma_\ell^2 e^{-\phi_\ell|t-s|}.
\]

We estimated the $\phi_\ell$ and $\sigma_\ell^2$ parameters 
using a least squares method on the observed sampled covariances, and 
then estimated $\boldsymbol\beta$ using
\begin{equation}
\label{beta_chapeuQMG}
\hat{ \boldsymbol \beta }= ( {\mathbf X}^\prime \hat{\boldsymbol\Sigma}^{-1} {\mathbf X} )^{-1}{\mathbf X}^\prime \hat{\boldsymbol\Sigma}^{-1} {\mathbf W},
\end{equation}

\noindent with ${\mathbf W}$, ${\mathbf X}$ as defined in 
(\ref{beta_chapeu}).

Now we replicate the experiment an additional 200 times, using the
same configuration of $\mathbf{y}$ for the calibration set, that is,
$y_{i,1}^\ast, y_{i,2}^\ast, y_{i,3}^\ast$ randomly generated with a
three-dimensional Dirichlet standard distribution. We repeated the
jackknife procedure and registered the estimated variances for the
$\hat{Y}_1^\ast$, $\hat{Y}_2^\ast$ and $\hat{Y}_3^\ast$ estimators,
for the $\phi = 0.5$ and $\phi = 0.002$ cases. The simulation study
compares several methods.  The functional approaches consist of basis
smoothing method when $K=14$ using ordinary least squares estimation
(OLS-K), basis smoothing method with generalized least squares
estimation proposed in \citeasnoun{dias:garcia:schmidt:2012} (GLS-K)
and Smoothing Splines method with tuning parameter $\lambda$
optimizing the GCV criteria (OLS-SS). For the classical multivariate
methods, we also offer jackknife estimates for Multiple Linear
Regression (MLR), Principal Components Regression using
either 3 components (PCR-o, from oracle, as the true calibration
curves have about three distinguishable features in the principal
components sense), or explaining 90\% of the sample variability of the
data (PCR-p, from proportion) and Partial Least Squares case (PLS-o, 
PLS-p, using the same number of components as in the Principal
Components Regression).

The median and interquartile range (IQR) results are given in Table
\ref{tab:sd2}. Notice that the theoretical value of all the standard
deviations is $\sqrt{1/18} \approx 0.24$ (since the values were
generated using a standard Dirichlet distribution). We can see that
increasing the number of sampled curves has little effect on the
precision of the estimators. On the other hand, in agreement with the
preliminary results shown in Table \ref{tab:sd}, having a strongly
correlated scenario for the sampling errors seems to produce less
precise estimators.  Our methods are overall better than Multiple
Linear Regression (MLR) but do not show the same degree of optimality
as the Principal Components and Partial Least Squares methods. We
emphasize that PCR/PLS methods, on the other hand, do not provide estimates
of the curves $\theta$ and are unable to separate the aggregated curves
for each analyte.

\begin{center}
[Table \ref{tab:sd2}  here]
\end{center}

To get estimates of both the bias and variability at prediction of new values, we ran a second experiment on the following $\mathbf{Y}^\ast$
\[
\mathbf{Y}^\ast = \left( \! \begin{array}{ccc}
 0.4 & 0.1 & 0.5 \\
 0.2 & 0.3 & 0.5 \\
 0.1 & 0.4 & 0.5 \\
 0.5 & 0.4 & 0.1
 \end{array} \! \right)
\]

First, we generated 40 independent learning sets, in which we
performed the calibration step using five models: K-basis spline with
ordinary least squares (OLS-K), smoothing spline with $\lambda$
minimizing GCV (OLS-SS), multiple linear regression (MLR), principal
components regression and partial least squares with a number of
components explaining $90\%$ of data variability (PCR-p and PLS-p,
respectively). For each of those 40 independent calibration sets, we
generated 5 independent curves $W_j^\ast(t)$ with concentrations based
on $\mathbf{Y}^\ast$ to be used for prediction purposes. Then, we
predict $\hat{Y}_1^\ast$, $\hat{Y}_2^\ast$ and
$\hat{Y}_3^\ast$. Indexing the 40 calibration sets with $g$ and the 5
replicates with $h$, we can estimate both the variability (V) and the
squared bias (B$^2$) using
\begin{eqnarray*}
\mbox{V}_c   & = & \sum_{g=1}^{40} \sum_{h=1}^5 (\hat{Y}_{gh,c} -
\bar{\hat{Y}}_{g \cdot ,c})^2 \\
\mbox{and} \\
\mbox{B}^2_c & = & \sum_{g=1}^{40} \sum_{h=1}^5 (Y_{gh,c} - \bar{\hat{Y}}_{g \cdot ,c})^2. \\
\end{eqnarray*}
The results of the prediction test for new, independent samples is
shown in Table \ref{tab:predictBiasVar}. One striking feature is how
much bias is introduced in all procedures when high correlation is
present.  In this case, the fixed K-basis and smoothing splines are
comparable; both perform better than Multiple Linear Regression but
have lower accuracy and precision than Principal Components Regression
and Partial Least Squares. Again, we stress the fact that these last
two methods do not offer estimates of the individual absorption
spectrum for each constituent, thus being better when prediction is
the sole interest of the experimenter. Furthermore, another
shortcoming of these multivariate procedures is that they cannot be
applied at all if the spectra are measured at different wavelengths.

\begin{center}
[Table \ref{tab:predictBiasVar} here]
\end{center}

\section{Real Datasets} \label{sec:data}

\subsection{PAH data}

The sample consists of 50 chemical samples of 10 polyaromatic
hydrocarbons (PAH) obtained by Electronic Absorption Spectroscopy,
each sample is composed of varying compositions of 10 different
constituents (pyrene, acenaphthene, anthracene, acenaphthylene,
chrysene, benzanthracene, fluoranthene, fluorene, naphthalene,
phenanthracene). Each sample was submitted to 27 wavelengths
(220nm--350nm). This dataset was presented by \citeasnoun{brer:2003}
to illustrate multivariate calibration and prediction techniques. 

This dataset is divided into two sets, 25 curves prepared to achieve
an orthogonal design with 5 levels of concentration for each
constituent to be used as a calibration sample. The other 25 curves
have the same five concentration levels for each constituent and can
be used for prediction purposes. One peculiarity of this dataset is
that the concentration is not given in percentage but in mg/l. This
has no effect in our estimation scheme. Notice that pyrene, chrysene, 
benzanthracene, fluorene and phenanthracene have generally higher 
concentrations than the other analytes.

\begin{figure}[!bpth]

\centering
\includegraphics[angle=-90,width=17cm,totalheight=20cm]{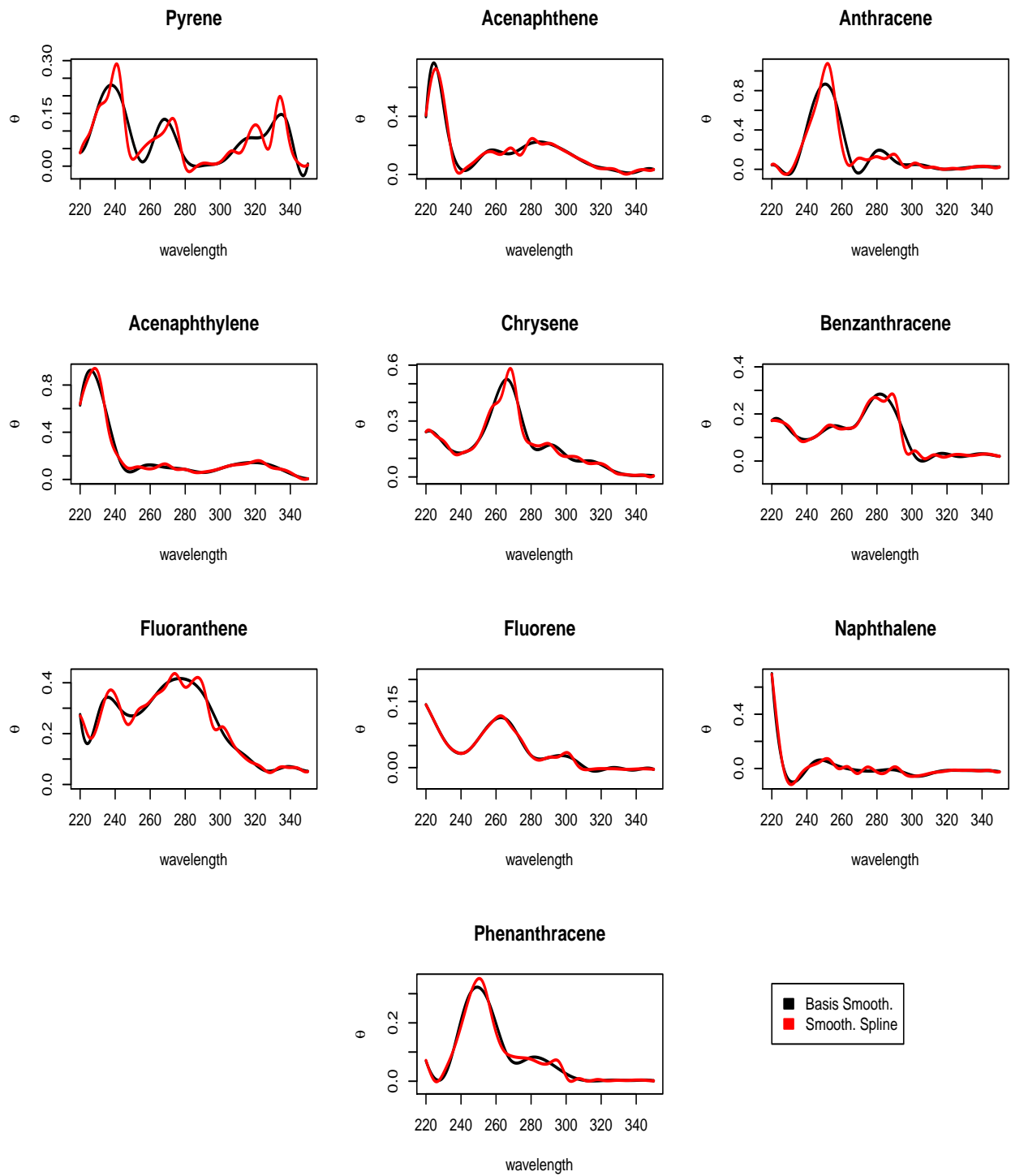}
\caption{Estimated absorbance spectra $\hat{\theta}_{\ell}$,
  $\ell=1,\ldots,10$ for PAH dataset}
\label{fig:pahTypology}
\end{figure}

\begin{figure}
	\centering
		\includegraphics[angle=-90,width=17cm]{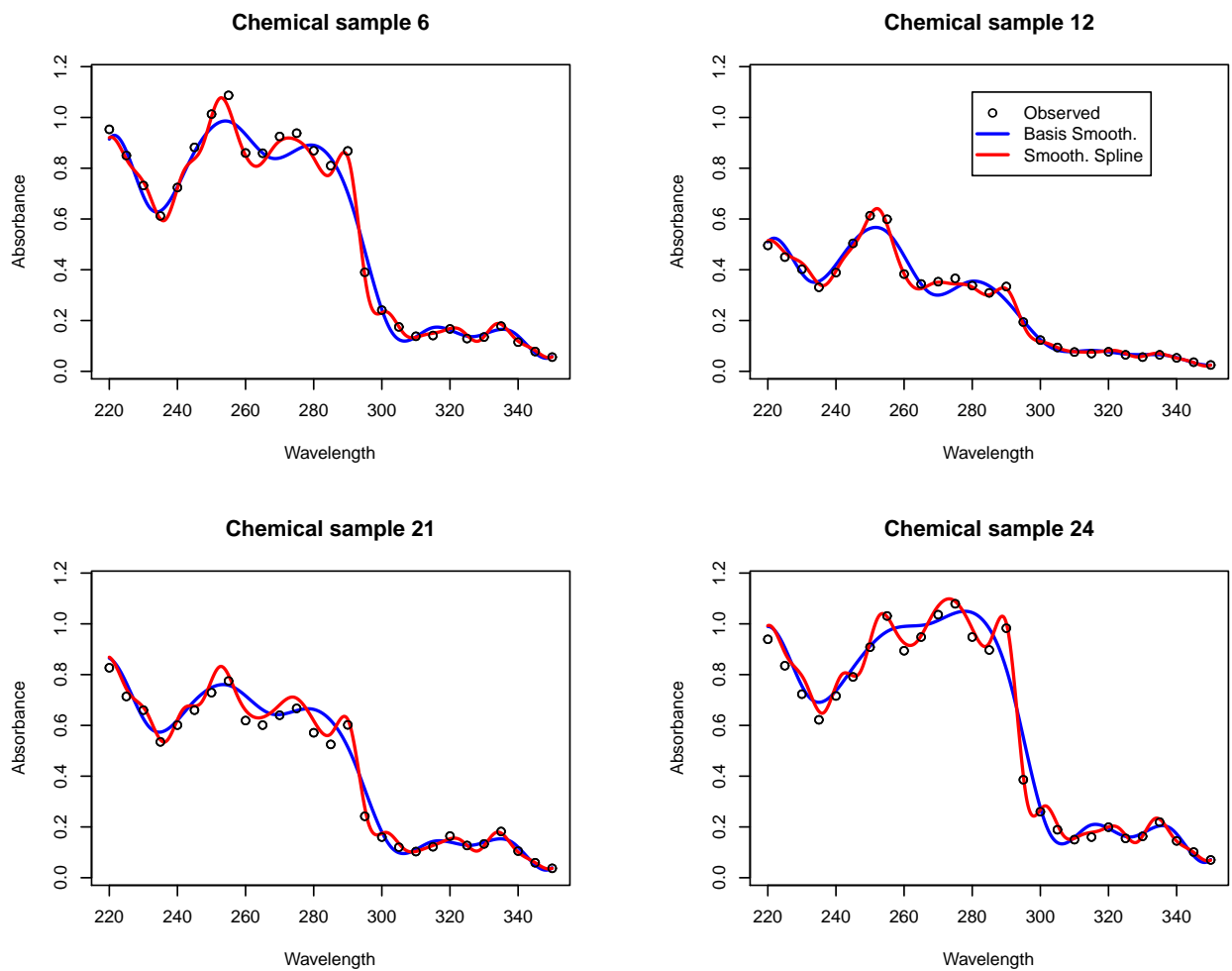} 
	\caption{Estimated aggregated curves ($\sum_{\ell=1}^m y_{i\ell}\hat{\theta}_{\ell}(t)$ for  chemical samples i=6,12, 21 and 24.}
	\label{fig:pahFittedSmoothB}
\end{figure}

Figure \ref{fig:pahTypology} compares the estimated analytes spectra
($\hat{\theta}_{\ell}(t), \ell=1,\ldots,10$) for the two proposed
basis expansion approaches. For this case, we used ordinary least
squares without any assumption on the covariance structure.  The
Smoothing Spline estimates are less smooth and have more ``bumps''.
Hence, it captures more of the local variation of the data. This is
expected and, for this dataset, a desirable property. As pointed by
\citeasnoun{brer:2003}, the curves are sampled at a coarse grid of
wavelengths and this causes the noise to be reduced. Therefore, most
of the local variation in the curves are important features of the
data and need to be captured. Figure \ref{fig:pahFittedSmoothB}
presents the fit for 4 chemical samples suggesting that our model
provides excellent fits to the observed aggregated data.

\begin{center}

[Figure \ref{fig:pahTypology} here]

[Figure \ref{fig:pahFittedSmoothB} here]

\end{center}

Notice that the fitted values using the Smoothing Splines are
much closer to the observed curves than the ones obtained via Basis
Smoothing. Finally, the leave-one-out technique for variance estimation 
of the predicted curves is employed, and the results are shown in
Table \ref{tab:LOUpah}. We remark that the estimated variances
are similar in both methods. 

\begin{center}
[Table \ref{tab:LOUpah} here]
\end{center}

\begin{figure}[!bpth]
\centering
\includegraphics[angle=-90,width=17cm,totalheight=20cm]{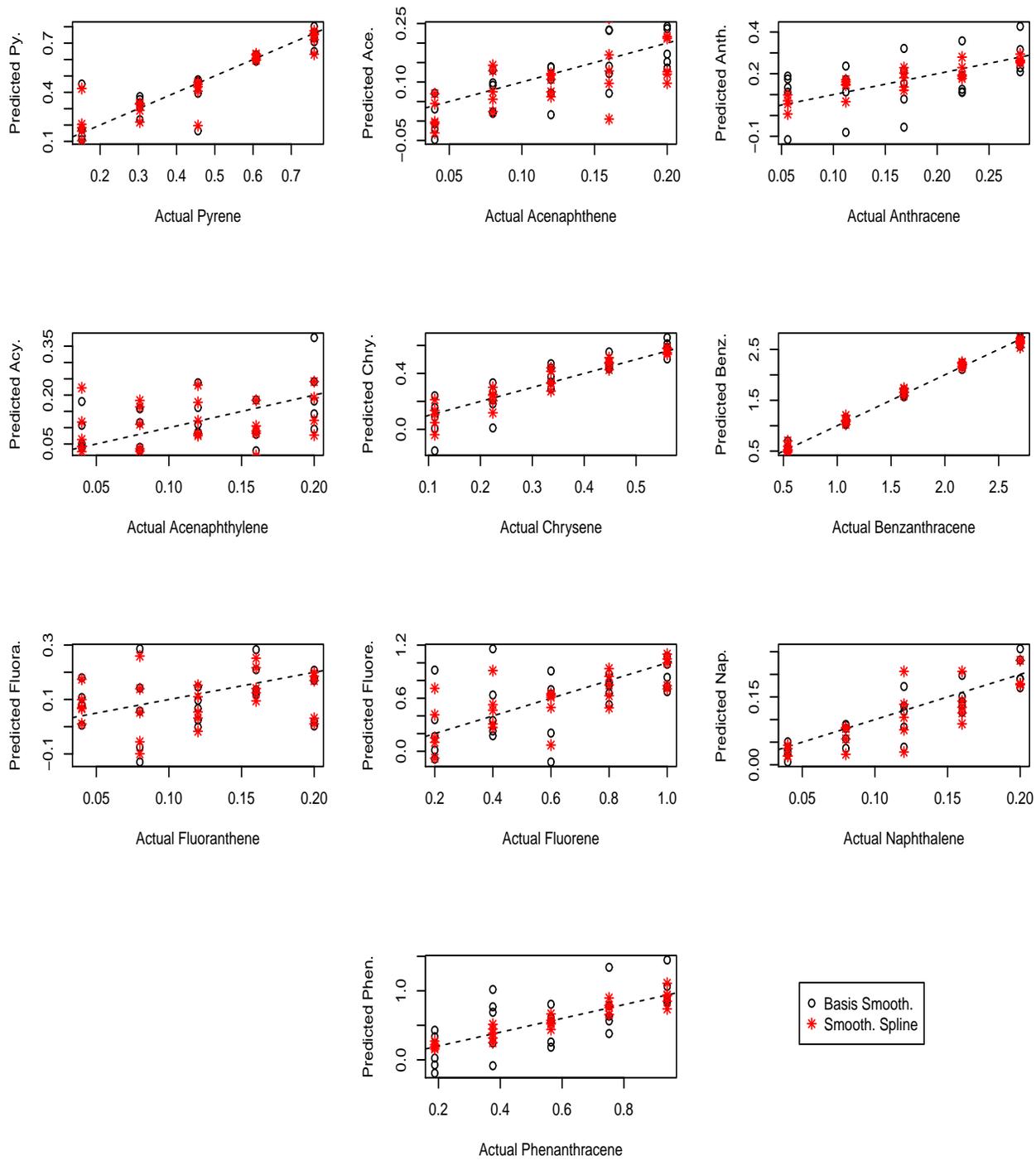}
\caption{The predicted values $\mathbf{y}$ for the Independent Test Set, for
  pyrene, acenaphthene, anthracene, acenaphthylene, chrysene,
  benzanthracene, fluoranthene, fluorene, naphthalene, phenanthracene
  and their predicted values.}
\label{fig:pahYpredicted}
\end{figure}

Of course, we aim not only for a fit with small residuals
but with high prediction power. Now Figure \ref{fig:pahYpredicted}
shows the $y_\ell^\ast$, $\ell=1,\ldots, 10$ for all the analytes for the
25 curves in the independent test set \cite[Table 5.20]{brer:2003},
against their predicted concentration for the model based on our
procedures. The dashed lines show where an exact prediction would be.
In general, both procedures have a similar performance in terms of
prediction. In fact, for benzanthracene, the constituent that
has the largest concentration for all the chemical samples (with
percentages ranging from 13.7\% to 59.6\%) both procedures have highly
predictive power. Other constituents that present
relatively high concentrations are: pyrene, chrysene, fluorene and
phenanthracene. We present the SEP (mg/l) for each component and the overall
SEP in Table \ref{tab:sep_PAH}, and again we provide a comparison with the 
MLR, PCR and PLS methods (we choose 10 components after Brereton, 2003).
We stress that the smoothing spline is capable of obtaining good 
predictions even when the respective concentrations are not 
particularly high, see for example anthracene. 

\begin{center}
[Table \ref{tab:sep_PAH} here]

[Figure \ref{fig:pahYpredicted} here]
\end{center}

\subsection{Tecator data}

These data are recorded on a Tecator Infratec Food and Feed analyzer
working in the wavelength range 850 -- 1050 nm by the Near Infrared
Transmission (NIT) principle. The data are made available as a
benchmark for regression models and it is available in the public
domain with no responsibility from the original data source from
Tecator\footnote{http://lib.stat.cmu.edu/datasets/tecator. The data
can be redistributed as long as this permission note is attached.} Each
sample contains finely chopped pure meat with different moisture, fat
and protein contents. The task is to predict the fat content of a meat
sample on the basis of its near infrared absorbance spectrum.

\begin{figure}[!bpth]
\centering
\includegraphics[angle=-90,width=17cm,totalheight=6cm]{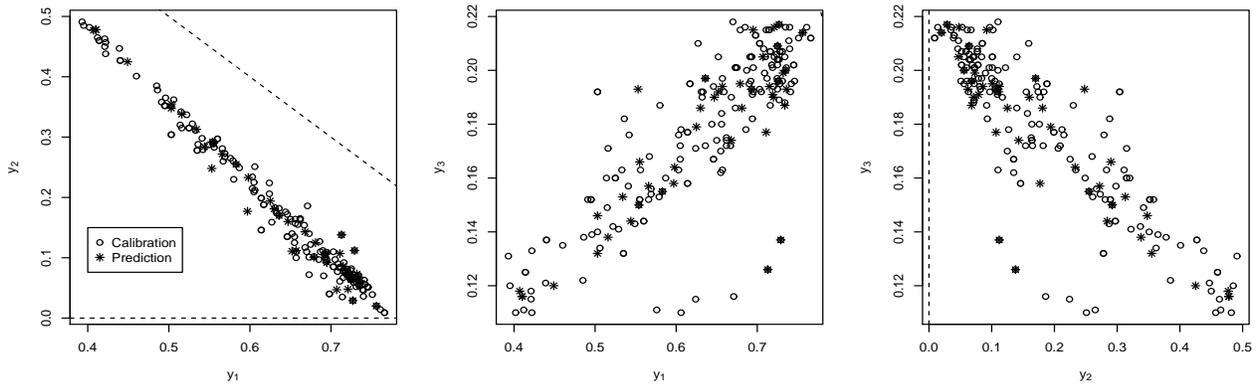}
\caption{Pairwise concentration values for moisture, fat and protein for the Tecator data}
\label{fig:tecatorConcentrastd}
\end{figure}

For each meat sample the data consists of a 100 channel spectrum of
absorbances and the contents of moisture (water), fat and protein. The
three contents, measured in percent, were determined by analytic
chemistry. Figure \ref{fig:tecatorConcentrastd} shows paired
scatterplots for the concentrations of moisture, fat and protein. It's
important to emphasize the regions for prediction of $\mathbf{y}$
values outside of these clouds will be, in some sense,
\emph{extrapolations} and consequently have very large prediction
errors.

\begin{center}
[Figure \ref{fig:tecatorConcentrastd}here]
\end{center}

There are 240 samples, further described by
\citeasnoun{borg:thod:1992}, divided into 5 data sets for the purpose
of model validation and extrapolation studies; the training set consisting
of 129 samples, the monitoring dataset consisting of 43 samples, the
testing dataset consisting of 43 samples, the extrapolation set for
fat content with 8 samples and the extrapolation set for protein
content with 17 samples.

The spectra are preprocessed using a principal component analysis on the 
data set C, and the first 22 principal components (scaled to unit variance) 
are included for each sample in the dataset. We did not, however, use this 
preprocessed data to calibrate our fit.

\begin{figure}[!b]
\centering
\includegraphics[angle=-90,width=17cm,totalheight=6cm]{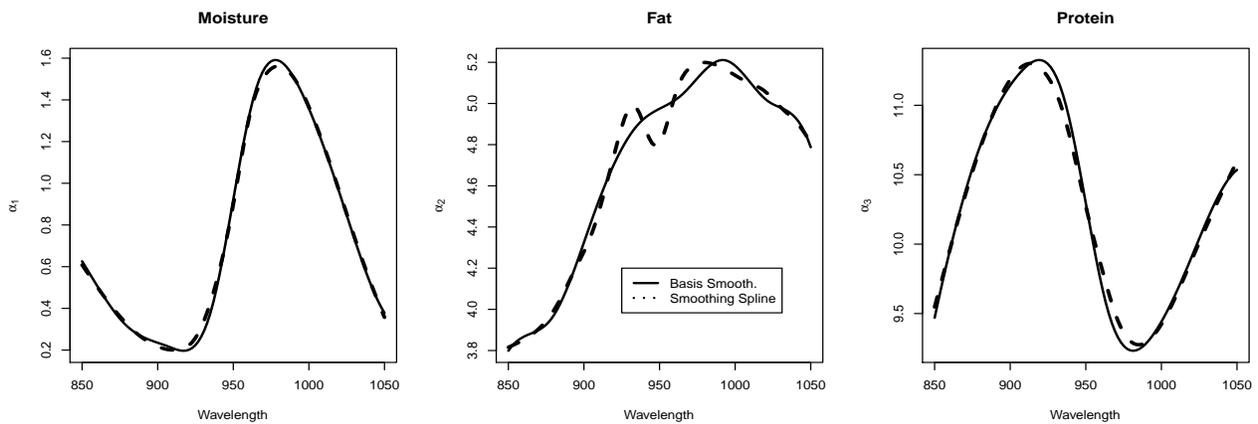}
\caption{Tecator estimated absorbance spectra $\hat{\theta}_1$, $\hat{\theta}_2$ and $\hat{\theta}_3$}
\label{fig:tecatorTypology}
\end{figure}

Figure \ref{fig:tecatorTypology} shows the absorbance spectra estimated with
the Calibration plus Monitoring sub-dataset for the Moisture, Fat and
Protein components. Notice here the smoothing spline seems to capture
some local features that are over smoothed when the number of basis is
kept fixed, particularly in the Fat component at wavelengths 920 to 990,
approximately. If we observe carefully the data, there is some important feature to
be captured in the curves at this part of the spectrum. 
However we choose not to use GCV to find $\lambda$, as the optimal
value provided by it ($\lambda = 33500$) oversmoothed the Protein
component and had consequently a worse performance in the prediction
step, in terms of SEP. We decided to pick $\lambda = 330$ by
inspecting plots of the estimated curves (like Figure \ref{fig:tecatorTypology}) until a satisfactory
degree of smoothness was found.

\begin{center}
[Figure \ref{fig:tecatorTypology} here]
\end{center}

Inspecting the residuals, we may see the errors probably fit the
strongly correlated data scenario. From the simulation conducted in
Section \ref{sec:simul} we expect the overall prediction error to be
greater than if we had a weakly correlated noise. 

\begin{figure}[!bpth]
 \centering
 \includegraphics[angle=-90,width=17cm,totalheight=6cm]{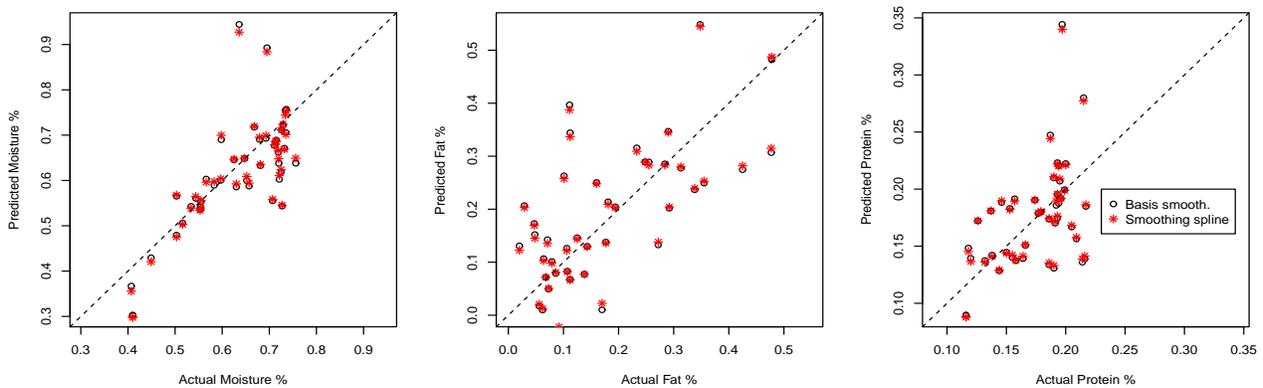}
 \caption{The $\mathbf{y}$ values for the Testing Data, for moisture, fat and protein, and their predicted values.}
 \label{fig:tecatorYpredicted}
 \end{figure}

Now Figure \ref{fig:tecatorYpredicted} shows the $y_1^\ast$,
$y_2^\ast$ and $y_3^\ast$ values of moisture, fat and protein --
respectively -- for the Testing data, against their predicted
concentrations for the model based on our procedure. Figure
\ref{fig:tecatorYpredicted} shows that both methods, fixed basis
smoothing and smoothing splines, have similar prediction power. The
dashed lines shows where an exact prediction would lie, so the given
standard deviation estimates (Table \ref{tab:sep2}) seem appropriate.

\begin{center}
 [Figure \ref{fig:tecatorYpredicted} here]

[Table \ref{tab:sep2} here]
\end{center}

Comparing the true values for the training data with their predicted 
correspondents, we find that the Standard Error of Prediction for the 
fat component on the prediction data set is equal to $0.102$ using
smoothing splines method. Table \ref{tab:sep}, originally taken from 
\citeasnoun{marx:etal:2009} but modified to include our results, show 
the SEP results for neural networks 
\citeaffixed{borg:thod:1992,thod:1995}{see}. For instance, 
our method is roughly 3.5 times more efficient than the 13-X-1 network 
in the prediction sense. \nocite{marx:etal:2009}

\begin{center}
[Table \ref{tab:sep} here]
\end{center}

\section{Concluding remarks}

In this work we propose to use a functional approach to analyze NIR
spectroscopy data. The main novelty of this work is to use a
functional version of the Beer-Lambert formula so that the calibration
procedure can be interpreted as modeling latent (disaggregated) mean
curves when we only have available observations of the population
(aggregated) curves. We considered two procedures for estimation of
the latent curves, basis smoothing and smoothing splines. In general,
smoothing splines capture more of the local variation of the data and
produce less smooth estimates. Once the calibration procedure is
performed, prediction was made by minimizing the squared error of
prediction. To assess the variance of such estimates, we used an idea
based on the leave-one-out jackknife technique.

To show the strength of the proposed functional approach, we analyzed
artificial datasets generated from the different correlation
structures and two examples with data available in the literature. We
compared our method with the most common methods available in the
literature, multiple linear regression (MLR), principal component
regression (PCR) and partial least squares (PLS). As a measure of fit
we used the Standard Error of Prediction (SEP) which is the root mean
squared difference between predictions and reference values. In all
cases, the fixed K-basis and smoothing splines are comparable; both
perform better than MLR but sometimes they have lower accuracy and
precision than the PCR and PLS. On the other hand, the functional
approach has two main advantages over these multivariate methods: i)
it can be applied if the spectra are measured at distinct wavelengths
for distinct curves; ii) it provides estimates for the individual
absorption curves for each constituent. Moreover, the analysis of the
artificial datasets demonstrates the impact of the temporal
correlation in the calibration and prediction steps showing the need
to model the correlation structure.

The applicability of our approach can also be viewed through the future
developments, already under work, which include: \\
(i) Implement Functional Principal Components or
other dimensionality reduction techniques to improve the performance
of the functional approach with respect to prediction only; \\ 
(ii) Perform jointly  estimation of the calibration curves and predicted concentration values; \\
(iii) Consider the case of outlier prediction in cases where the prediction sample is inconsistent with the calibration data.

\paragraph{Acknowledgments} The authors would like to thank Prof. John
Rice, U. C. Berkeley for many fruitful discussions. This work was
partially funded by CNPq grants 302755/2010-1, 476764/2010-6,
302182/2010-1 and 553438/2009-3. The authors are grateful to
Department of Mathematics of the University of Wisconsin-Madison and
Department of Statistics of the University of California, Berkeley,
for their hospitality.  

\bibliography{myref}
\pagebreak

\begin{table}
	\centering
		\begin{tabular}{c|cc|cc} \hline
 & \multicolumn{2}{|c|}{Weakly Correlated} & \multicolumn{2}{|c}{Strongly Correlated} \\ \hline
       & $I=20$ & $I=100$ & $I=20$ & $I=100$ \\ \hline
 $S_1$ & 0.020  & 0.016   & 0.061  & 0.061   \\
 $S_2$ & 0.027  & 0.024   & 0.042  & 0.055   \\
 $S_3$ & 0.022  & 0.021   & 0.058  & 0.063   \\ \hline
		\end{tabular}
	\caption{Estimated standard deviations for the $\hat{y}^\ast_1$, $\hat{y}^\ast_2$ and $\hat{y}^\ast_3$, using the jackknife procedure}
	\label{tab:sd} 
\end{table}

\pagebreak

\begin{table}[!bpth]
  \footnotesize
	\centering
		\begin{tabular}{l|l|l|ccc|ccccc}
		\hline
		$\phi$                 & I                      & C      & OLS-K     & GLS-K     & OLS-SS    & MLR       & PCR-o    & PCR-p     & PLS-o     & PLS-p     \\
		\hline
		\multirow{8}{*}{0.5}   & \multirow{3}{*}{20}    & 1      & 1.8 (0.4) &           & 1.9 (0.5) & 2.0 (0.4) & 1.9 (0.4) & 1.9 (0.5) & 2.0 (0.5) & 2.0 (0.6) \\
		                       &                        & 2      & 2.5 (0.6) &           & 2.6 (0.6) & 2.8 (0.6) & 2.9 (0.6) & 2.9 (0.8) & 3.0 (0.8) & 3.1 (1.0) \\
		                       &                        & 3      & 2.1 (0.5) &           & 2.1 (0.5) & 2.2 (0.5) & 1.8 (0.4) & 1.8 (0.5) & 1.8 (0.4) & 1.9 (0.6) \\
		\cline{3-11}           
		                       &                        & $\ast$ & \textbf{2.1 (0.7)} &           & \textbf{2.2 (0.7)} & \textbf{2.3 (0.7)} & \textbf{2.0 (0.9)} & \textbf{2.1 (0.9)} & \textbf{2.1 (1.0)} & \textbf{2.2 (1.0)} \\
		\cline{2-11}
		                       & \multirow{3}{*}{100}   & 1      & 1.7 (0.2) &           & 1.7 (0.2) & 1.8 (0.2) & 1.6 (0.2) & 1.9 (0.2) & 1.7 (0.2) & 3.7 (0.9) \\
		                       &                        & 2      & 2.3 (0.2) &           & 2.3 (0.2) & 2.3 (0.2) & 2.3 (0.2) & 2.7 (0.4) & 2.6 (0.3) & 5.3 (1.5) \\
		                       &                        & 3      & 1.8 (0.2) &           & 1.9 (0.2) & 1.9 (0.2) & 1.5 (0.2) & 1.8 (0.2) & 1.6 (0.2) & 3.6 (0.8) \\
		\cline{3-11}           
		                       &                        & $\ast$ & \textbf{1.9 (0.5)} &           & \textbf{1.9 (0.5) }& \textbf{1.9 (0.5)} & \textbf{1.7 (0.7)} & \textbf{2.0 (0.8)} & \textbf{1.8 (0.8)} & \textbf{4.0 (1.5)} \\
		\hline
		\multirow{8}{*}{0.002} & \multirow{3}{*}{20}    & 1      & 6.7 (1.5) & 6.6 (1.5) & 6.9 (1.6) & 6.9 (1.6) & 4.3 (1.0) & 4.3 (1.0) & 4.3 (1.0) & 4.3 (1.0) \\
		                       &                        & 2      & 5.4 (1.5) & 5.4 (1.4) & 5.6 (1.6) & 5.6 (1.6) & 4.3 (1.0) & 4.3 (1.0) & 4.3 (1.0) & 4.3 (1.0) \\
		                       &                        & 3      & 7.5 (2.1) & 7.3 (2.0) & 7.7 (2.2) & 7.7 (2.2) & 3.7 (0.9) & 3.7 (0.9) & 3.7 (0.9) & 3.7 (0.9) \\
		\cline{3-11}           
		                       &                        & $\ast$ & \textbf{6.5 (2.0)} & \textbf{6.5 (1.9)} & \textbf{6.7 (2.0)} & \textbf{6.7 (2.0)} & \textbf{4.0 (1.1)} &\textbf{ 4.1 (1.1)} & \textbf{4.1 (1.1)} & \textbf{4.1 (1.1)} \\
		\cline{2-11}
		                       & \multirow{3}{*}{100}   & 1      & 6.1 (0.6) & 6.1 (0.6) & 6.1 (0.7) & 6.1 (0.7) & 3.7 (0.4) & 3.6 (0.3) & 3.7 (0.4) & 3.7 (0.4) \\
		                       &                        & 2      & 5.0 (0.5) & 5.1 (0.6) & 5.1 (0.5) & 5.1 (0.5) & 3.8 (0.4) & 3.8 (0.4) & 3.8 (0.4) & 4.0 (0.5) \\
		                       &                        & 3      & 7.0 (0.6) & 6.9 (0.7) & 7.0 (0.7) & 7.0 (0.7) & 3.3 (0.3) & 3.2 (0.3) & 3.3 (0.3) & 3.3 (0.4) \\
		\cline{3-11}           
		                       &                        & $\ast$ & \textbf{6.1 (1.5)} & \textbf{6.1 (1.5)} & \textbf{6.1 (1.5)} & \textbf{6.1 (1.5)} & \textbf{3.6 (0.5)} & \textbf{3.6 (0.5) }& \textbf{3.6 (0.5)} & \textbf{3.6 (0.6)} \\
		\hline
		\end{tabular}
                \caption{$10^2$ times Median ($10^2$ times IQR) of 200 independent standard deviation simulation estimates for the $\hat{Y}_1^\ast$, $\hat{Y}_2^\ast$ and $\hat{Y}_3^\ast$ estimators, using the jackknife procedure. OLS-K: fixed 14 spline basis with ordinary least squares; GLS-K: fixed 14 spline basis with generalized least squares (only for $\phi = 0.002$); OLS-SS: smoothing spline ($\lambda$ minimizing GCV); MLR: multiple linear regression; PCR-o, PLS-p: Principal Components Regression with either 3 components selected or with $p$
                  components that explain $90\%$ of observed curves
                  variability; PLS-o, PLS-p: Partial Least Squares, same
                  number of components as corresponding PCR. True
                  standard deviation = $2.4 \times 10^{-2}$}
	\label{tab:sd2} 
\end{table}

\begin{table}[ht]
\centering
\begin{tabular}{r|rrrr|rrrrrr}
  \hline
                      & \multicolumn{2}{c}{OLS-K} & \multicolumn{2}{c|}{OLS-SS} & \multicolumn{2}{c}{MLR} & \multicolumn{2}{c}{PCR} & \multicolumn{2}{c}{PLS} \\ 
  \hline              & B$^2$ &  V   & B$^2$ &  V   & B$^2$ &  V   & B$^2$ &  V   & B$^2$ &  V   \\
  \hline I = 20       & 0.13  & 0.38 & 0.14  & 0.37 & 0.14  & 0.36 & 0.13  & 0.34 & 0.14  & 0.40 \\ 
  I = 20, correlated  & 1.33  & 3.26 & 1.33  & 3.26 & 1.36  & 3.32 & 0.45  & 1.26 & 0.45  & 1.26 \\ 
  I = 100             & 0.08  & 0.30 & 0.08  & 0.30 & 0.08  & 0.30 & 0.10  & 0.33 & 0.42  & 1.35 \\ 
  I = 100, correlated & 0.79  & 2.81 & 0.79  & 2.81 & 0.80  & 2.89 & 0.23  & 0.97 & 0.24  & 1.03 \\ 
   \hline
\end{tabular}
\caption{$10^3$ times estimated squared bias and variability of
  the prediction set, for all components added up: OLS-K:
  fixed K=14 spline basis with ordinary least squares;
  OLS-SS: smoothing spline with $\lambda$ minimizing GCV; MLR: multiple linear regression; PCR: Principal Components Regression with $p$ components that explain $90\%$ of observed curves variability; PLS: Partial Least Squares, same number of components as PCR.}
	\label{tab:predictBiasVar} 
\end{table}

\begin{table}
\centering 
\begin{tabular}{l|c|c} \hline
Constituent & Basis Smoothing (OLS-K) & Smoothing Splines (OLS-SS) \\ \hline
Pyrene &  0.06 &   0.05 \\	
Acenaphthene &  0.04 &   0.07 \\	
Anthracene  & 0.11  &  0.04 \\	
Acenaphthylene  & 0.41  &  0.42 \\	
Chrysene  & 0.28  &  0.27 \\	
Benzanthracene  & 1.64  &  1.64 \\	
Fluoranthene  & 0.41  &  0.41 \\	
Fluorene  & 0.59  &  0.56 \\	
Naphthalene  & 0.11  &  0.11 \\	
Phenanthracene  & 0.49  &  0.36 \\ \hline	
\end{tabular}
\label{tab:LOUpah}
\caption{Leave-one-out Standard Deviation estimates (mg/l) for PAH data.}
\end{table}

\begin{table}
\centering 
\begin{tabular}{l|c|c|c|c|c} \hline
Constituent     & OLS-K & OLS-SS & MLR  & PCR  & PLS  \\
\hline Pyrene   & 0.09  &  0.09  & 0.09 & 0.09 & 0.09 \\	
Acenaphthene    & 0.05  &  0.06  & 0.06 & 0.03 & 0.03 \\	
Anthracene      & 0.11  &  0.04  & 0.03 & 0.03 & 0.03 \\	
Acenaphthylene  & 0.08  &  0.09  & 0.09 & 0.06 & 0.06 \\	
Chrysene        & 0.10  &  0.06  & 0.06 & 0.05 & 0.04 \\	
Benzanthracene  & 0.06  &  0.07  & 0.07 & 0.07 & 0.06 \\	
Fluoranthene    & 0.10  &  0.09  & 0.09 & 0.08 & 0.08 \\	
Fluorene        & 0.32  &  0.23  & 0.24 & 0.20 & 0.15 \\	
Naphthalene     & 0.03  &  0.04  & 0.04 & 0.03 & 0.04 \\	
Phenanthracene  & 0.31  &  0.09  & 0.08 & 0.09 & 0.08 \\ 
\hline Overall  & 0.16  &  0.10  & 0.11 & 0.09 & 0.08 \\ \hline
\end{tabular}
\caption{Standard Error of Prediction (mg/l) for PAH data. OLS-K: fixed-K (14) spline basis with ordinary least squares; OLS-SS: smoothing spline ($\lambda$ minimizing GCV); MLR: multiple linear regression; PCR: Principal Components Regression with 10 components (Brereton, 2003); PLS: Partial Least Squares, same number of components as PCR.}
\label{tab:sep_PAH}
\end{table}

\begin{table}
\centering
\begin{tabular}{c|cc|cc} \hline
 & \multicolumn{2}{|c|}{OLS-K} & \multicolumn{2}{|c}{OLS-SS} \\ \hline
 Component $\ell$ & $S_\ell$ & $\mathrm{SEP}_\ell$ & $S_\ell$ & $\mathrm{SEP}_\ell$ \\ \hline
 Moisture & 0.386 & 0.081 & 0.386 & 0.080 \\
 Fat      & 0.514 & 0.106 & 0.513 & 0.102 \\
 Protein  & 0.159 & 0.040 & 0.159 & 0.039 \\
 \hline
 Overall  &       & 0.081 &       & 0.077 \\
 \hline
\end{tabular}
\caption{Prediction $S_\ell$ estimates and SEP for individual
  components. OLS-K: fixed K=14 spline basis with ordinary least
  squares; OLS-SS: smoothing spline with $\lambda$ minimizing GCV.} 
\label{tab:sep2}
\end{table}

\begin{table}
\centering
\begin{tabular}{c|c} \hline
 Method & SEP \\ \hline
10-6-1 network, early stopping & 0.650 \\
10-3-1 network, Bayesian & 0.520 \\
13-X-1 network, Bayesian, Automatic Relevance Determination & 0.360 \\
Basis Smoothing & 0.106 \\
Smoothing Spline & 0.102 \\ \hline
\end{tabular}
\caption{Table summarizing the prediction error for fat in the Tecator data, compared to a neural network approach; first three entries are tabulated data available with Tecator dataset.}  
\label{tab:sep}
\end{table}

\pagebreak

\pagebreak

\pagebreak

\end{document}